\documentclass[showpacs,aps,twocolumn
]{revtex4}
\usepackage{graphicx}
\usepackage{amsfonts}
\usepackage{amsmath}
\usepackage{amssymb}
\usepackage{multirow}
\usepackage{braket}
\usepackage[usenames,dvipsnames]{color}
\usepackage[english]{babel}
\usepackage[cp1251]{inputenc}   

\begin{document}
\bibliographystyle{apsrev}

\title{Information entropy and  temperature of the binary Markov chains}

\author{O.~V.~Usatenko$^{1}$, S.~S.~Melnyk$^{1}$, G.~M.~Pritula$^{1}$, and V.~A. Yampol'skii$^{1,2}$}

\affiliation{$^1$ A.~Ya. Usikov Institute for Radiophysics and Electronics NASU, 61085 Kharkov, Ukraine\\
$^2$ V.~N. Karazin Kharkov National University, 61077 Kharkov, Ukraine}

\begin{abstract}We propose two different approaches for introducing the information
temperature of the binary $N$-th order Markov chains.
The first approach is based on a comparison of the Markov sequences with the equilibrium Ising chains  at
given temperatures. The second approach uses probabilities of
finite-length subsequences of symbols occurring, which determine
their entropies. The derivative of the entropy with respect to the
energy gives the information temperature measured
on the scale of introduced energy. For the case of nearest neighbor spin/symbol interaction, both approaches give similar results. However, the
method based on the correspondence of the $N$-step Markov and Ising chains appears to be very
cumbersome for $N>3$. We also introduce the information temperature for the weakly correlated one-parametric Markov chains and present results for the
step-wise and power memory functions. An application of the developed method to obtain the information temperature of some literary texts is given.
\end{abstract}
\maketitle

\section{Introduction}

In statistical mechanics, the axiomatic introduction of the
entropy as logarithm of the statistical weight is one of the possible
ways for constructing statistical thermodynamics of a
micro-canonical ensemble for arbitrary physical system. The next step
in this direction consists in establishing the condition of
equilibrium between two weakly interacting systems. The requirement
of the maximum number of states for the entire system leads to the
equality of the derivatives of the entropies $S$ with respect to the
energy $E$ for both subsystems. This allows us to introduce the value $T$, called
the temperature, $1/T=dS/dE$.

C. Shannon~\cite{Shan} has generalized the notion of entropy to
other systems, in particular, to numerical and symbolic random
sequences. This entropy is introduced in a way similar to the ordinary thermodynamic one. It is often written (see,
e.g., Refs.~\cite{Cover,Ebeling}) as the entropy $H_{L}$ of subsequence of the length $L$ of an infinite random stationary ergodic
sequence $\mathbb{S}$ of symbols $a_{i}$ belonging to the alphabet
$\mathcal{A}$,
\begin{eqnarray} \label{entro_block}
H_{L}=-\sum_{a_{1},...,a_{L} \in \mathcal{A}} P(a_{1}^{L})\log_{2}
P(a_{1}^{L}).
\end{eqnarray}
Here $P(a_{1}^{L}) =P(a_{1},\ldots,a_{L})$ is the probability to
find the $L$-word $a_{1}^{L}$ in the sequence.

Definition~\eqref{entro_block} does not contain the energy of
symbols (or letters or words in the text), so the temperature
can not be introduced in the ordinary thermodynamic way. Nevertheless,
B. Mandelbrot \cite{Mandel} proposed the idea of the possibility and
usefulness of introducing the concept of information temperature for
texts considered as random sequences. The temperature measurement of
a text in a textbook would allow one to determine its academic level from
its vocabulary and could serve as a measurement of the vocabulary
complexity of books in general. Afterward  the concept of ``text
temperature'' was applied to linguistic analysis of the texts
\cite{Campos,Kosmidis,Rego,Chang} under the assumption that human
language could be described as a physical system within the framework
of equilibrium statistical mechanics. This approach was used to
analyze the high-frequency words with the use of the Boltzmann
distribution~\cite{Miyazima} and, in a different way, to consider
the low-frequency vocabulary \cite{Rovenchak}.
It is difficult to
accept the presupposition about the statistical independence of words
in the text used nearly in all cited works. Nevertheless, the
following simple qualitative considerations indicates possibility of
the information temperature introduction for different cases.

Actually, we can imagine two limiting cases of ordering of symbols in
a random sequence. In a weakly correlated binary sequence of
elements ``zero'' and ``one'', these elements are barely ordered and
this sequence can be regarded as a chain of symbols disordered by
high temperature. A strongly correlated random sequence, governed by
the conditional probability distribution function of the ordinary
Markov chain with one-step conditional probability function
$P(1|1)=P(0|0)=1/2+\mu$ in the limiting case $ \mu \rightarrow 1/2$
contains large blocs of ``zeros'' and ``unities'', which can be
considered as the attraction of the same type symbols at low
(with respect to the energy interaction proportional to $\mu$)
temperature.

We can consider an even simpler example of a Markov random sequence of
non-interacting symbols generated with probability $P(1)=1/2 + \nu$
for ``1'' and $P(0)= 1/2 - \nu$ for ``0''. Now, if we look at this
sequence as at a 1D collection of noninteracting spin particles in a
magnetic field, then, after identification of symbols and spin
particles, $1\Leftrightarrow\,\uparrow$,
 $0\Leftrightarrow\,\downarrow$, and using the Boltzmann distribution,
we can easily find a correspondence between the parameter $\nu$ and
the ratio of the magnetic field to the temperature:
\begin{equation}\label{nu}
\nu=\frac{1}{2}\tanh(\mathcal{H}/T)\quad \,\Leftrightarrow\,\quad \frac{1}{\tau} =
\frac{1}{2} \ln \frac{1+2\nu}{1-2\nu}.
\end{equation}
So, just as the temperature $T$ defines the spin distribution in the Ising chain, how it looks like at high and low temperatures at given $H$, the parameter $\tau$ controls the symbols distribution in the Markov random sequence. Thus, we can consider the parameter $\tau$ as the effective temperature of the Markov chain.

Relying on this qualitative reasoning, we develop here
two quantitative approaches for introducing the information
temperature of binary random sequences. One of these approaches is based on the fact that every binary Markov chain is equivalent to some binary two-sided sequence \cite{AMUYa} which, in its turn, can be viewed as the Ising chain where the probability of a configuration is given by the Boltzmann distribution at fixed temperature.

The second method is based on the thermodynamic definition of temperature and uses direct calculations of the block entropy and energy of a Markov chain in the pair-interaction  approximation. The blocks are chosen to be $(N+1)$-length subsequences of symbols since they encompass all the information about the structure of the $N$-step Markov chain. The probabilities of the blocks/words are governed by the Chapman-Kolmogorov equation (see Section~\ref{ChKEq}).  Solutions of this equation  can be hardly found for a general case of the $N$-th order chain, so we use an exactly solvable models to calculate the entropy. The derivative of the latter with respect to the energy gives the inverse information temperature. Both the approaches give similar results.

The structure of this paper is as follows. In Section \ref{GenDef}, we provide a brief description of models, definitions, and previous results
necessary for understanding the further presentation. In Section
\ref{LowOd}, we introduce the information
temperature for the binary $N$-th order Markov chains with
step-wise memory functions for $N\leqslant 3$.
We first apply both our approaches to analyze the simplest, ordinary or one-step,
Markov chain and show that they lead to the same result for the information temperature. Then we
repeat the procedure for the second-order chains and
present the expression for entropy-based temperature
in the case of the third-order one-parametric Markov chain. In Section
\ref{WeekCorr}, we obtain evaluations of the information temperatures
for the $N$-th order weakly correlated Markov chains with step-wise
and power memory functions. In the end of this Section, we apply our approach to define the information temperature for some literary texts. The last
Section \ref{Discussion} contains conclusion and discussion of the
prospects for the further researches.
\section{General definitions}\label{GenDef}
%
Here we introduce the definitions of the binary symbolic Markov
chain with step-wise conditional probability distribution function
(CPDF) and discuss the equivalence of the Markov chain and two-sided random
sequences and the property of entropy defined by Eq.~\eqref{entro_block}.

\subsubsection{Symbolic Markov chains}
 Consider a semi-infinite random stationary ergodic sequence
$\mathbb{S}$  of symbols-numbers $a_{i}$,
\begin{equation}
\label{RanSeq} \mathbb{S}= a_{0}, a_{1},a_{2},...
\end{equation}
taken from the binary alphabet
\begin{equation}\label{alph}
 \mathcal{A}=\{0,1\},\,\, a_{i}\in \mathcal{A},\,\, i \in
\mathbb{N}_{+} = \{0,1,2...\}.
\end{equation}
We use the notation $a_i$ to indicate a position $i$ of the symbol
$a$ in the chain and the unified notation $\alpha^k$ to stress the
value of the symbol $a\in \mathcal{A}$. 

We suppose that the symbolic sequence $\mathbb{S}$ is a
\textit{high-order Markov chain}. The sequence $\mathbb{S}$ is a
Markov chain if it possesses the following property: the probability
of symbol~$a_i$ to have a certain value $\alpha^k \in \mathcal{A}$
under the condition that {\emph{all}} previous symbols are fixed
depends on $N$ previous symbols only,
\begin{eqnarray}\label{def_mark}
&& P(a_i=\alpha^k|\ldots,a_{i-2},a_{i-1})\\[6pt]
&&=P(a_i=\alpha^k|a_{i-N},\ldots,a_{i-2},a_{i-1}).\nonumber
\end{eqnarray}

There are many other terms for such sequences. They are also
referred to as: \emph{categorical}~\cite{Hoss},
\textit{higher-order}~\cite{Raftery, Seifert}, \emph{multi-} or
$N$-\emph{step}~\cite{UYa,RewUAMM} Markov's chains. One of the most
important and interesting application of the symbolic sequences is
the probabilistic language model, which specializes in predicting
the next item in a sequence by means of $N$ previous known symbols.
In such applications, the Markov chains are known as the $N$-\emph{gram models}.

A particular form of the CPDF,
$P(a_i=\alpha^k|a_{i-N},\ldots,a_{i-2},a_{i-1})$, of the binary
$N$-step Markov chain is \cite{muya},
\begin{equation}\label{CondPr_power}
  P\left( a_i=1 | a_{i-N}^{i-1} \right)
    =  \overline{a} + \sum_{r=1}^{N} F(r) (a_{i-r}-\overline{a} ),
\end{equation}
where we use the concise notation $a_1^{N}=a_1,a_2,...,a_{N}$. We
refer to such sequences as the additive Markov chains and to $F(r)$
as the \emph{memory function} (MF). It describes the strength of
influence of previous symbol $a_{i-r} (1 \leqslant r \leqslant N$)
upon a generated one, $a_{i}$.

\subsubsection{ Markov's chain with step-wise memory function}\label{ChKEq}
In Refs.~\cite{UYa} and~\cite{UYaKM}, the \emph{step-wise}  model of Markov chain was introduced. In this model, the conditional probability $p_{k}$ of
occurring the symbol ``0'' after the $N$-symbol words containing $k$
unities, e.g., after the word
$(\underbrace{11...1}_{k}\;\underbrace{00...0}_{N-k})$, is given by
the following expression:
\[
p_{k}=P(a_{N+1}=0\mid \underbrace{11\dots
1}_{k}\;\underbrace{00\dots 0} _{N-k})
\]
\begin{equation}
=\frac{1}{2}+\mu (1-\frac{2k}{N}).  \label{1}
\end{equation}

If we focus our attention on the region of $\mu $ determined by the
persistence inequality, $0 < \mu <1/2$, then each of the
symbols unity in the preceding $N$-word promotes the birth of new
symbol unity. The region $-1/2<\mu <0$ corresponds to the anti-persistent (anti-ferromagnetic) symbols-spins ordering. At $\mu \rightarrow 0$, the random chain becomes a completely
disordered white noise, for $\mu \rightarrow \pm 1/2 $, we have
strongly correlated sequences consisting of large blocks of\, ``0''\,
and ``1''\, at $\mu \rightarrow 1/2 $, and containing long
anti-ferromagnetic blocks with perfectly alternating symbols\, ``0''
\, and\, ``1''\, at $\mu \rightarrow - 1/2 $ .

\subsubsection{Chapman-Kolmogorov equation}
For the stationary Markov chain, the probability $b(a_{1}a_{2}\dots
a_{N})$ of occurring a certain word $(a_{1},a_{2},\dots ,a_{N})$
satisfies the condition of compatibility for the Chapman-Kolmogorov
equation (see, for example, Ref.~\cite{gar}):
\[
b(a_{1}\dots a_{N})
\]
\begin{equation}
=\sum_{a=0,1}b(aa_{1}\dots a_{N-1})P(a_{N}\mid a,a_{1},\dots
,a_{N-1}).  \label{10}
\end{equation}
Here we have $2^{N}$ homogeneous algebraic equations for
$2^{N}$ probabilities $b$ of occurring the $N$-words and the
normalization equation $\sum b=1$. In the case under consideration,
the set of equations can be substantially simplified owing to its
following property.

For the Markov chain with step-wise memory function, the probability
$b(a_{1}a_{2}\dots a_{N})$ depends only on the number $k$ of unities in
the $N$-symbol word, i.\ e., it is independent of the
arrangement of symbols in the word $(a_{1},a_{2},\dots ,a_{N})$.
This yields the recursion relation for
$b(k)=b(\underbrace{11...1}_{k}\; \underbrace{00...0}_{N-k})$,
\begin{equation}
b(k)=\frac{q_{k-1}}{p_{k}}b(k-1),  \label{15}
\end{equation}
where $q_{k}=1-p_{k}$ is the conditional probability of occurring the
symbol ``1'' after the $N$-word containing $k$ unities.

\subsubsection{Equivalence of Markov and two-sided
sequences}\label{2sided}
The probability that symbol~$a_i$ is equal to unity, under condition
that the \emph{rest} of symbols in the random chain are fixed, can
be presented in the form~\cite{AMUYa},
\begin{equation} \label{2}
P(a_i=1|A_i^-,A_i^+)
\end{equation}
\begin{equation}
=\displaystyle\frac{P(a_i=1,A_i^+|A_i^-)}
{P(a_i=1,A_i^+|A_i^-)+P(a_i=0,A_i^+|A_i^-)},\nonumber \\[10pt]
\end{equation}
where $A_i^-$ and $A_i^+$ are the semi-infinite words,
$(\ldots,a_{i-2},a_{i-1})$ and $(a_{i+1},a_{i+2},\ldots)$,
surrounding symbol $a_i$, respectively, are
\begin{equation}
\underbrace{\ldots,a_{i-2},a_{i-1}}_{A_i^-},a_i,\underbrace{a_{i+1},a_{i+2},\ldots}_{A_i^+}.
\end{equation}
Here the two-sided probability $P(a_i=1|A_i^-,A_i^+)$ is expressed
by means of the Markov-like probability functions $P(a_i
,A_i^+|A_i^-)$. These probabilities containing semi-infinite
one-sided words can be expressed in terms of the conditional
probability function of the Markov chain. For this, one needs to use
iteratively Eq.~(\ref{def_mark}) in order to present
$P(\cdot|\cdot)$ in a factorized form,
\begin{widetext}
\[
P(a_i=1,A_i^+|A_i^-)=P(a_i=1|A_i^-)
P(a_{i+1},A_{i+1}^+|a_i=1,A_i^-)=
 \]
\[
=P(a_i=1|A_i^-)P(a_{i+1}|a_i=1,A_i^-)
P(a_{i+2},A_{i+2}^+|A_i^-,a_i=1,a_{i+1})=\ldots
 \]
\begin{equation}\label{equiv_dsv2}
\ldots=\prod \limits_{{r=0 \atop a_i=1}}^{N}
P(a_{i+r}|A_{i+r}^-)\prod\limits_{{r=N+1\atop a_i=1}}^{\infty}
P(a_{i+r}|A_{i+r}^-).
\end{equation}
\end{widetext}

Thus, Eqs.~\eqref{2} and \eqref{equiv_dsv2} specify the correspondence
between two-sided random sequences and ordinary (one-sided) Markov
chains. In fact, we will consider the Markov chain of finite order
(with finite memory). By this reason, the sign of infinity in
Eq.~\eqref{equiv_dsv2} have to be replaced by the Markov
chain order $N$.
\subsubsection{Conditional entropy }
Equation~\eqref{entro_block} makes it possible to introduce the
\emph{conditional} entropy, which is the entropy per symbol,
\begin{eqnarray} \label{ShennEntr}
h_{L}= H_{L+1} - H_{L}.
\end{eqnarray}
This value specifies the degree of uncertainty of the $(L + 1)$-th
symbol observation if the preceding $L$ symbols are known. The
source entropy is the entropy per symbol at the asymptotic limit, $h
= \lim_{L\rightarrow \infty}
 h_L$.  It measures the average information per symbol if all correlations, in the
statistical sense, are taken into account. The conditional entropy
$h_L$ can be presented (see details in Appendix A) in terms of the
conditional probability distribution function (CPDF),
Eq.~\eqref{def_mark},

\begin{eqnarray} \label{ShennCondEntr}
h_{L}= \overline{ h(a_{L+1}|a_{1}^{L})},
\end{eqnarray}
\begin{eqnarray}\label{Entro_cond}
   h(a_{L+1}|a_{1}^{L}) = - \!\!\sum_{a_{L+1} \in \mathcal{A}}\!\!
P(a_{L+1}|a_{1}^{L})\log_2 P(a_{L+1}|a_{1}^{L}),
    \label{siL}
\end{eqnarray}
where the line over $h$, $\overline{h(.|.)}$, denotes a statistical
average. The function $P(a_{L+1}|a_{1}^{L})$ is
$(a_{k})$-independent at $k\leqslant 0$. By this reason, $h_{L}$ is
constant at $L\geqslant N$ (see Appendix A),
\begin{eqnarray} \label{SEN}
h_{L>N}= h_{N}.
\end{eqnarray}
%
\subsubsection{Ising chain of classical spins}\label{IsingCh}
We consider the \emph{chain of classical spins} with Hamiltonian
\begin{equation}\label{hamilt}
\mathrm{H}=-\sum\limits_{j-i \leqslant N\atop i<j}\varepsilon(j-i)s_i s_j,
\end{equation}
where $s_i$ is the spin variable taking on two values, $-1$ and $1$,
and $\varepsilon(r) = \varepsilon(-r)$ is the exchange integral of
the coupling. The range $N$ of spin interaction may be arbitrary, but
finite. We will suppose below that the range of spin particle
interaction is always equal to the order of the corresponding Markov chain.

The binary variables $a_i=\{0,1\}$ of the Markov chain is related
to the spin Ising's variables $s_i=\{-1,1\} =
\{\downarrow,\uparrow\}$ by the equality
\begin{equation}
s_i = 2 a_i - 1.
\end{equation}
%
\section{Information Temperature of low-order Markov chains}\label{LowOd}

In this section, we present two ways to introduce the
information temperature for the Markov chains of the orders $N \leqslant 3$
with the step-wise conditional probability function \eqref{1} and demonstrate
their equivalence.

\subsection{Temperature of the ordinary Markov chains, $ N=1$}

First, we introduce the information temperature for the simplest
case of the one-step Markov chain using two approaches.
\subsubsection{Equivalence/correspondence approach}
Considering the equivalence of Markov chain to the
two-sided random sequence, we apply
Eq.~\eqref{2} to the word $(a_{i-1},1,a_{i+1})$:
\begin{eqnarray}\label{TwoOne}
&& P(a_i=1|a_{i-1},a_{i+1})\\[6pt]
&&=\frac{P(a_i=1,a_{i+1}|a_{i-1})}{P(a_i=1,a_{i+1}|a_{i-1})+
P(a_i=0,a_{i+1}|a_{i-1})}.\nonumber
\end{eqnarray}
For the ordinary Markov chain with property~\eqref{1},
we easily obtain the conditional probabilities
\begin{eqnarray}\label{111}
&& P(a_i=1,a_{i+1}=1|a_{i-1}=1)=\left(\frac{1}{2}+\mu\right)^2,\\[6pt]
&&P(a_i=0,a_{i+1}=1|a_{i-1}=1)=\left(\frac{1}{2}-\mu\right)^2,\nonumber
\end{eqnarray}
and the Boltzmann distribution for the corresponding Ising model,
\begin{eqnarray}\label{P Izing}
&& P(a_i=1|a_{i-1}=1,a_{i+1}=1)\nonumber \\[6pt]
&&=\frac{\exp(2\varepsilon/T)}{\exp(2\varepsilon/T)+\exp(-
2\varepsilon/T)}.
\end{eqnarray}
The energies of the spins interaction are $\varepsilon_{\uparrow
\uparrow }=\varepsilon_{\downarrow\downarrow}=-\varepsilon$,
$\varepsilon_{\uparrow \downarrow
}=\varepsilon_{\downarrow\uparrow}=\varepsilon>0.$ On the base of
Eqs.~\eqref{TwoOne} and \eqref{P Izing}, the
relation between the information temperature and the correlation
parameter $\mu$ can be found as follows:
\begin{equation}
  \frac{\left( \frac{1}{2}+\mu \right)^2}
            {{\left( \frac{1}{2}+\mu \right)^2 +
            {\left( \frac{1}{2}-\mu \right)^2}}}
        =
        \frac{\exp(2\varepsilon/T)}{\exp (2\varepsilon/T) + \exp(-2\varepsilon/T)},
 \label{eq:MuT}
\end{equation}
Note that for the word (1,1,0), which
is equivalent to the word (0,1,1) in our case, the relation between
the information temperature and the correlation parameter is
reduced to the identity.

Solution of Eq.~\eqref{eq:MuT} gives
\begin{equation}
  \mu = \frac{1}{2} \tanh {\left( \frac{1}{\tau} \right)}, \quad\quad
  \frac{1}{\tau} = \frac{1}{2} \ln \frac{1+2\mu}{1-2\mu}.
 \label{MuVsT}
\end{equation}
So, for $\tau \rightarrow {\pm\infty}$, we have $\varepsilon/T \simeq 2\mu
\rightarrow 0$, and $\tau \rightarrow \pm 0$ when $\mu \rightarrow
\pm {1/2}$.

\begin{figure}[h!]
\begin{centering}
\scalebox{0.6}[0.8]{\includegraphics{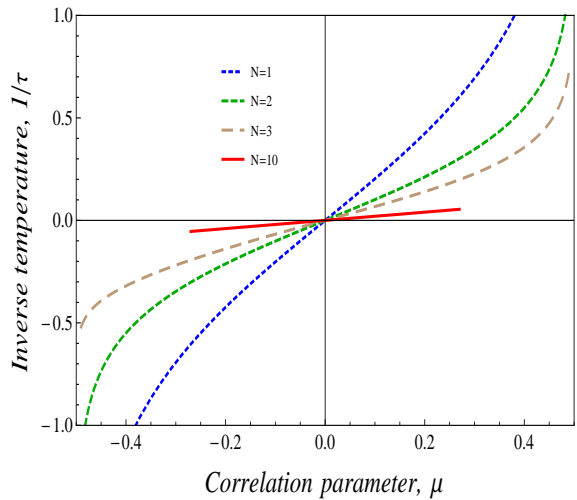}}
\caption{The dependence $\tau ^{-1}(\mu)$ for $N$-step Markov chains with step-wise memory functions for $N=1,\,2, 3$ (the corresponding lines are marked in the legend), and for the weakly correlated Markov chain with $N=10$. These plots are obtained within the entropy approach.} \label{Fig1}
\end{centering}
\end{figure}

Here we have introduced the information temperature $\tau=
T/\varepsilon$ of the ordinary (one-step) Markov chain and $\varepsilon/T$
is the dimensionless inverse information temperature. The negative values of $\tau$
describe an anti-ferromagnetic ordering of spines or symbols\,
``0''\, and ``1''. We can say that $\tau$ is the temperature $T$
measured in unites of  energy $\varepsilon$, $\tau={T}/\varepsilon$. The result of our consideration for the dependence of
$\tau ^{-1}$ on the parameter $\mu$ is presented by the short-dashed
curves in Fig.~\ref{Fig1}.

\subsubsection{Entropy based temperature}\label{2a}
In this subsection, we introduce the information temperature of the one-step Markov sequences by means of its definition via entropy and energy, $1/T=dS/dE$. Similarly to the first approach, we also compare the Markov sequence with an Ising chain. However, here we do not care that the chain of spins obeys the Boltzmann distribution at a given temperature, but just arrange mentally the spins ``up'' and ``down'' at the places of the Markov chain where the symbols ``one'' and ``zero'' stand, respectively. Then we calculate the energy and entropy of 3-symbol fragments of the Ising chain and introduce the inverse temperature as the derivation of the entropy over energy.

Using the probabilities of the 2-symbol words occurring,
\begin{equation}\label{00}
P_{00}=P_{11}=(1/2+\mu)/2,
\end{equation}
\begin{equation}\label{01}
    P_{01}=P_{10}=(1/2-\mu)/2,
\end{equation}
we derive the average fictive energy per one bond of symbols-spins,
\begin{equation}
  E_2 = \sum_{i,j \in \{0,1\} }\varepsilon_{ij} P_{ij}=-2\varepsilon
  \mu,
\end{equation}
and the entropy per one bond
\begin{equation}
  H_2 = -2 P_{00}\ln P_{00}-2 P_{01}\ln P_{01}= \ln 2 \nonumber
  \end{equation}
\begin{equation}\label{S}
 -(1/2+\mu) \ln (1/2+\mu)-(1/2-\mu) \ln (1/2-\mu).
\end{equation}
Calculating the derivation $dH_2/dE_2$ as $(dH_2/d\mu)/(dE_2/d\mu)$, we obtain
the same result as is presented by Eq.~\eqref{MuVsT}, i.e., two
different methods of calculation turn out to be consistent. Here, a
natural question arises: what will we get in the case of more
distant symbol interaction, for the two-step Markov chains?
\subsection{Two-step Markov chains, $N=2$}\label{EandE}
\subsubsection{Equivalent/correspondence approach}
Consider the second order Markov chain \eqref{1} and the
corresponding Ising chain with  the range of interaction $N=2$. In
this case, Eq.~\eqref{2} takes the form:
\begin{eqnarray}\label{TwoStep}
&& P(a_i=1|A^{-},A^{+})\\[6pt]
&&=\frac{P(a_i=1,A^{+}|A^{-})}{P(a_i=1,A^{+}|A^{-})+P(a_i=0,A^{+}|A^{-})},\nonumber
\end{eqnarray}
where $A^{\pm}=a_{i\pm 1},a_{i\pm 2}$. We suppose that energies of
the central spin interaction with the nearest neighbors in the Ising
model are $\varepsilon_{\uparrow \uparrow
}=\varepsilon_{\downarrow\downarrow}=-\varepsilon_1$,
$\varepsilon_{\uparrow \downarrow
}=\varepsilon_{\downarrow\uparrow}=\varepsilon_1$ and next to the
nearest energies of interaction are $\varepsilon_{\uparrow \bullet
\uparrow }=\varepsilon_{\downarrow \bullet
\downarrow}=-\varepsilon_2$, $\varepsilon_{\uparrow \bullet
\downarrow }=\varepsilon_{\downarrow \bullet
\uparrow}=\varepsilon_2.$

Considering the probabilities of occurrence of different configurations of symbols, we obtain from
Eqs.~\eqref{P Izing} and \eqref{TwoStep} two independent equations,
%
\begin{equation}
  \frac{1}{2}+\mu=
            \frac{\exp(2\varepsilon_2/T)} {\exp (2\varepsilon_2/T) + \exp(-2\varepsilon_2/T)},
 \label{eq:MuT_1}
\end{equation}
\begin{equation}
  \frac{\left( 1+2\mu \right)^2}
            {\left(1 +2\mu \right)^2 + 1}
        =
        \frac{\exp(2\varepsilon_1/T)}{\exp (2\varepsilon_1/T) + \exp(-2\varepsilon_1/T)}.
 \label{eq:MuT_2}
\end{equation}
Solutions of these equations express the ratio of energies $\varepsilon_1$ and $\varepsilon_2$,
\begin{equation}
  \frac{\varepsilon_1}{\varepsilon_2} = \frac{1}{2} \frac{\ln[(1+2\mu)/(1-2\mu)]}{\ln (1+2\mu)},
  \label{y1}
\end{equation}
and the normalized to $\varepsilon_1$ and/or to $\varepsilon_2$ temperature $T$ of the Ising chain,
\begin{equation}
  \frac{T}{\varepsilon_1}= \frac{2}{\ln (1+2\mu)}, \qquad
  \frac{T}{\varepsilon_2}= \frac{4}{\ln [(1+2\mu)/(1-2\mu)]},
\label{y2}
\end{equation}
via the correlation parameter $\mu$ of the equivalent Markov sequence. The obtained parameters of the Ising chain satisfy the following
relationship:
\begin{equation}
  \exp\left(2\varepsilon_1/T\right)
    \left[1+\exp\left(-4\varepsilon_2/T\right)  \right]=2.
 \label{eq:restr}
\end{equation}

Note that, for fixed value of potential $\varepsilon_2$, the energy $\varepsilon_1$ tends to zero at $\mu \rightarrow 1/2$. Therefore, the normalization of information temperature to $\varepsilon_1$ is not suitable, and we define the dimensionless temperature $\tau$ as
\begin{equation}
 \tau= \frac{T}{\varepsilon_2}=\frac{4}{\ln [(1+2\mu)/(1-2\mu)]}.
 \label{y4}
\end{equation}

%
\subsubsection{Energy and entropy of 3-symbol words}\label{3a}
In this subsection, we use the second (entropy-based) approach for introducing the information temperature for the two-step Markov sequences.

Using Eqs.~\eqref{1} and \eqref{15} for $N=2$, we find the
probabilities to meet 2-symbols words,
\begin{equation}\label{002}
    P_{00}=P_{11}=\frac{1}{4(1-\mu)},
\end{equation}
\begin{equation}\label{012}
    P_{01}=P_{10}=\frac{1-2\mu}{4(1-\mu)}.
\end{equation}
These probabilities differ from those given by Eqs.~\eqref{00} and
\eqref{01} for the one-step Markov chain.

The 3-point probabilities are
\begin{equation}\label{000}
    P_{000}=P_{111}=\frac{1+2\mu}{8(1-\mu)},
\end{equation}
\begin{equation}\label{001100}
    P_{001}= ... = P_{110}=\frac{1-2\mu}{8(1-\mu)}.
\end{equation}
Energies of different spin configurations are:
\begin{equation}\label{000111}
    \varepsilon(\downarrow\downarrow\downarrow)=\varepsilon(\uparrow\uparrow\uparrow)
    =-2\varepsilon_1-\varepsilon_2,
\end{equation}
\begin{equation}\label{010101}
    \varepsilon(\downarrow \uparrow \downarrow)=\varepsilon(\uparrow \downarrow\uparrow)
    =2\varepsilon_1-\varepsilon_2,
\end{equation}
and
\begin{equation}\label{allothers}
    \varepsilon(\downarrow \downarrow \uparrow)=...
    =\varepsilon(\uparrow \uparrow\downarrow)=\varepsilon_2.
\end{equation}
Equations~\eqref{000}~-~\eqref{allothers} allow us to calculate the
averaged energy and entropy per two bonds of random elements,
\begin{equation}\label{2Temperature}
    E_3 = -(2\varepsilon_1+\varepsilon_2)\frac{\mu}{1-\mu},
\end{equation}

\begin{equation}\label{2S}
  H_3 = -2 P_{000}\ln P_{000}-6 P_{001}\ln P_{001}.
  \end{equation}
Using Eqs.~\eqref{000} and \eqref{001100}, after differentiation, we
have
\begin{equation}\label{d2S}
  \frac{\partial H_3}{\partial \mu} = \frac{3}{4(1-\mu)^2}\ln\frac{1-2\mu}{1+2\mu}.
  \end{equation}
Calculating derivative $dH/dE$ as $(dH_3/d\mu)/(dE_3 /d\mu)$, we
obtain the following result:
\begin{equation}\label{2tau}
    \frac{1}{\tau}=\frac{1}{4} \ln
    \frac{1+2\mu}{1-2\mu},
    \end{equation}
which formally coincides with Eq.~\eqref{y4}. However, here we use another normalization coefficient,
\begin{equation}\label{AvEps}
   \tau=\frac{T}{\langle\, \varepsilon\rangle }, \quad\, \langle\, \varepsilon\rangle=
    \frac{2\,\varepsilon_1+\varepsilon_2}{3}.
\end{equation}
The value $\langle\, \varepsilon\rangle $ can be considered as a peculiar average of the energy of spin interaction per one adjacent spin in a 3-symbol fragment of the Ising chain. Of course, this energy differs from the true average energy $E_3$ given by Eq.~\eqref{2Temperature}.
The important point of
the above consideration is independence of function
$\tau (\mu)$ from the introduced arbitrary energies
$\varepsilon_1$ and $\varepsilon_2$ of the symbols interactions.

The limiting property of above obtained expression for $\tau$ is
\begin{equation}\label{lim2tau}
    \varepsilon/T=\frac{1}{\tau}\simeq\mu \quad \texttt{at} \quad \mu\rightarrow 0.
\end{equation}
The result \eqref{2tau} is depicted in Fig.~\ref{Fig1} by the green dashed curve.
%
\subsubsection{Inverse problem}

It is of interest to solve the inverse problem of defining a Markov sequence with true two-step memory function which provide the same statistical properties as the equilibrium Ising chain with given  one-step and two-step potentials $\varepsilon_1$ and $\varepsilon_2$ independent of each other. We assume that the Ising chain obeys the Boltzmann statistics. The desired Markov chain is given by the conditional probability function of
more general form in comparison with that given by Eq.~\eqref{1},
\begin{equation}
P(a_3|a_1,a_2)=\frac{1}{2}+(-1)^{a_3}\!\!\left[\mu_1
\!\!\left(\frac{1}{2}\!-\!a_2\right) \!\!+\!\!\mu_2
\!\!\left(\frac{1}{2}\!-\!a_1\right)\right]. \label{eq:P}
\end{equation}
The joint probability for the two-sided chain in terms of interaction energies
$\varepsilon_1$ and $\varepsilon_2$ is
\begin{equation}
P(a,\!b,\!1,\!c,\!d)=\! \frac{{B_2}^{a+d-1} {B_1}^{b+c-1}}{\!{B_2}^{-(a+d-1)}
\!{B_1}^{-(b+c-1)}+\!{B_2}^{a+d-1} \!{B_1}^{b+c-1}},
 \label{eq:}
\end{equation}
where $B_{1,2}=\exp{\left(2\varepsilon_{1,2}/T\right)}$.

The equations which relate the energies $\varepsilon_1$ and
$\varepsilon_2$ to the correlation constants $\mu_1$ and
$\mu_2$ are
\begin{equation}
\exp{\left(2\varepsilon_{1}/T\right)}=
\frac{1+{\mu_2}+{\mu_1}}{1+{\mu_2}-{\mu_1}},
 \label{eq:bet1}
\end{equation}
and
\begin{equation}
\exp{\left(4\varepsilon_{2}/T\right)}=
\frac{(1+\mu_2)^2-\mu_1^2}{(1-\mu_2)^2-\mu_1^2}\,.
 \label{eq:bet2}
\end{equation}
The information temperature normalized to $\varepsilon_2$ is given by equation,
\begin{equation}
\tau=\frac{T}{\varepsilon_2} =
4\ln^{-1} \frac{(1+\mu_2)^2-\mu_1^2}{(1-\mu_2)^2-\mu_1^2}\,.
 \label{tau_2}
\end{equation}
This equation is reduced to Eq.~\eqref{2tau} at $\mu_1= \mu_2=\mu$. 

If the energies $\varepsilon_1$ and $\varepsilon_2$ are subjected to restriction Eq.~\eqref{eq:restr}, model Eq.~\eqref{eq:P} becomes one-parametric with $\mu_1=\mu_2=\mu$.

Now we can compare two suggested methods for introducing the information temperature of the $N$-th step Markov chains, namely, the equivalence method and the entropy one. The first method, being  more correct and better physically justified, is more capricious since it is not always possible to find a suitable Ising model with Boltzmann distribution statistically equivalent to the $N$-th order Markov chain.  In addition, for $N>2$, this method becomes very cumbersome.

\subsection{Entropy based temperature for the 3-rd order sequences, $N=3$} \label{n3}

Here we present the results of calculations of the information temperature for 3-step Markov chains, performed within the entropy based approach. These calculations are similar to that made in previous subsection. The details are given in Appendix B. We give the expression for the dimensionless inverse temperature in the form,
%
\begin{eqnarray}
\frac{1}{\tau}&=&\frac{1}{72}\left[\left(-4 \mu ^2+6 \mu +9\right)
\log \left(\frac{1+2 \mu }{1-2 \mu}\right)+\right. \nonumber \\
&&\left. 3\left(4 \mu ^2-6 \mu +3\right) \log \left(\frac{3+2 \mu }{3-2 \mu }\right)\right],
\label{eq:9}
\end{eqnarray}
where we use the definitions,
\begin{equation}
\frac{1}{\tau}=\frac{\partial{H_4}/\partial{\mu}}{\partial{E_4
}/\partial{\mu}}= \frac{ \langle\, \varepsilon\rangle }{T},
\,\,\,\quad \langle\, \varepsilon\rangle =\frac{3 \varepsilon _1+2
\varepsilon _2+\varepsilon _3}{6}.
\label{eq:51}
\end{equation}
At $\mu \rightarrow 0$ the value of $1/\tau $ is
\begin{equation}\label{tau3}
   1/\tau \approx 2\mu/3.
\end{equation}
Note, that with increase of $N$ the calculations become more and more
cumbersome, and it is impossible to obtain a general result for
information temperature at large values of $N$. However, supposing that the
correlations in random chain are weak, we can find the desired
results. We can move further in the study of the temperature of the
Markov chain if we assume that the random sequence is weakly
correlated, when the memory function in Eq.~\eqref{CondPr_power} satisfies  the condition %
\begin{equation}\label{Eq:WeakCorr}
   \sum_{r=1}^{N}| F(r)| \ll 1.
\end{equation}
%

\section{Temperature of additive Markov chains with weak
correlations}\label{WeekCorr}
\subsection{ $N$-step Markov chain with step-wise
memory function}

In this section, we study the weakly correlated  Markov sequences. First we consider the chains
with weak step-wise memory function.

\subsubsection{Entropy }
To calculate the block entropy of the $N$-step  Markov chain with
step-wise memory function \eqref{1} we can use the results
for the conditional entropy $h_L$ obtained in Refs.~\cite{MelUs,UYaKM}. For
the case of weak correlations, it has the form (see Ref.~\cite{MelUs})
\begin{equation}\label{Eq:EntropDiff}
    h_L = h_0 - \frac{1}{2} \sum_{r=1}^L K^2(r),
\end{equation}
where $h_0=\ln 2$, $K(r)$ is the correlation coefficient, $K(r) =
C(r)/C(0)$ where
$C(r)=\overline{(a_i-\overline{a})(a_{i+r}-\overline{a})}$ is the
correlation function of the chain. For $L \leqslant N$, we have
$K(r)=1/(1+2n)$ with the  persistence parameter $n = N(1-2\mu)/4\mu$
(see, e.g., Ref.~\cite{UYaKM}). So, in this case, the conditional entropy Eq.~\eqref{ShennEntr} per symbol for the block of $L$-length is
\begin{equation} \label{hl}
h_L = \ln 2 - \frac{L}{2(1+2n)^2}, \quad L \leqslant N.
\end{equation}
Thus, we can calculate the entropy for the block of length
$N+1$,
\begin{equation}\label{Eq:HN}
H_{N+1} = \sum_{r=0}^{N} h_r = (N+1) \left(\ln 2 -
\frac{N}{4(1+2n)^2} \right).
\end{equation}%
In the limiting case of weak persistence,  $n \gg 1$, we have
\begin{equation}\label{Eq:limHN}
    H_{N+1} = (N+1) \left(\ln 2 - \frac{\mu^2}{N} \right).
\end{equation}

\subsubsection{Energy}
Let us now calculate the average energy of the block  of length
$N+1$,
\begin{equation} \label{AvBlockE}
    E_{N+1} = \left< E(a_1^{N+1}) \right>
        = \sum_{w_{N+1}} P(a_1^{N+1}) E(a_1^{N+1}),
\end{equation}
where the averaging is done over all symbols of the word
$w_{_{N+1}} =a_1^{N+1}=a_1,a_2,...,a_{N+1}$.

Considering these words as $(a_1^{N},1)$ and $(a_1^{N},0)$, we can
rewrite Eq.~\eqref{AvBlockE} in more detailed form,

\begin{equation}\label{Eq:EwN1}
    E_{N+1}
        = \sum_{w_N} \left( P(a_1^{N},0)
    E(a_1^{N},0) + P(a_1^{N},1)E(a_1^{N},1) \right).
\end{equation}
For the Markov chain model with step-wise memory function Eq.~\eqref{def_mark}, we can
replace the summation over words $w_N$ by the summation over number
of units $k$ in these words, and present the probabilities of the
words $(a_1^{N},1)$ and $(a_1^{N},0)$ as
\begin{equation}\label{Eq:Ew3}
    P(a_1^{N},0) = W_{N}(k)
                \left( \frac{1}{2} -\mu (2k/N - 1) \right),
\end{equation}
\begin{equation}\label{Eq:Ew4}
    P(a_1^{N}, 1) = W_{N}(k)
        \left( \frac{1}{2} +\mu (2k/N - 1) \right).
\end{equation}
Here $W_{N}(k)=\left(\begin{array}{c}N\\k\end{array}\right)b(k)$ is
the probability that the word $a_1^{N}$ contains $k$ unities with
arbitrary order of symbols, $b(k)$ is defined by Eq.~\eqref{15}, and
$\left(\begin{array}{c} N \\ k \end{array} \right)$ is the number of
possible $k$-combinations.

Assume that all the pair-interaction energies in the corresponding Ising chain depend on the spin orientations but do not depend on the distance between them,
$\varepsilon_{n}(a_i, a_j)=-\varepsilon (2a_i - 1) (2a_j - 1)$.
Then, for the total energy of $a_1^{N+1}$ word, we have
\begin{equation}
    E(a_1^{N+1}) = -\frac{\varepsilon}{2} \sum_{i \neq j} (2a_i - 1) (2a_j - 1),
\end{equation}
or, in terms of $k$,
\begin{equation}\label{Eq:Ew0}
    E(a_1^{N+1}) = - \frac{\varepsilon}{2} \left[ (2k' - N - 1)^2  - N - 1 \right],
\end{equation}
where $k' = k$ for the word $(a_1^{N},0)$ and  $k' = k+1$ for the
word $(a_1^{N},1)$.

Using Eqs.~\eqref{Eq:Ew3}, \eqref{Eq:Ew4}, and  \eqref{Eq:Ew0},
we obtain from Eq.~\eqref{Eq:EwN1} the average energy,
\begin{equation}
    E_{N+1} = -\frac{\varepsilon}{2} \left[ 4D_N - N + \frac{8\mu}{N}
    \sum_{k=0}^N W_N(k)(2k-N)^2 \right],
\end{equation}
where $D_N=\overline{k^2}-{\overline{k}}^2$ is the variance, the
averaging is done over all $N$-words.

For the case of weak persistence, $n \gg 1$, the distribution
$W_N(k)$ is a narrow Gaussian function with the variance
$D_N=N/4(1-\mu)$ (see, e.g., Ref.~ \cite{UYaKM}),
\begin{equation}
    W_N(k) = \frac{1}{\sqrt{2 \pi D_N}} \exp\left(
-\frac{(k-N/2)^2}{2D_N} \right).
\end{equation}
Supposing $N\gg 1$, we can change the summation by integration and
obtain
\begin{equation}\label{Eq:E}
        E_{N+1} = -\frac{\varepsilon \mu}{1-2\mu}\left( N+1 \right)
        \approx -(N+1) \mu \varepsilon.
\end{equation}
\subsubsection{Temperature}
Using Eqs.~\eqref{Eq:limHN} and \eqref{Eq:E}, we arrive at the
following evaluation for the information temperature:
\begin{equation}\label{muN}
   \frac{1}{\tau} = \frac{\varepsilon}{T} = \varepsilon \frac{dH_{N+1}}{d\mu}
   \Big/ \frac{dE_{N+1}}{d\mu} = \frac{2 \mu}{N}.
\end{equation}
This result surprisingly coincides with
Eqs.~\eqref{MuVsT}, \eqref{2tau}, and \eqref{eq:9} for the limiting cases of $N=1,2,3$,
despite the fact that upon obtaining the result
Eq.~\eqref{muN} we assumed $N \gg 1$. It is very probable that Eq.~\eqref{muN} is valid  not only at $N \gg 1$ but for any $N$.
\subsection{Markov chain with long-range power memory function}

Here we consider the binary $N$-step  Markov chain defined by the
additive conditional probability function Eq.~\eqref{CondPr_power} with
the power-law memory,
\begin{equation}\label{Eq:pow}
    F(r) = F_0 r^{-\alpha}.
\end{equation}

The energy of interaction between two elements in the Ising chain is also
supposed to be the power function,
\begin{equation}\label{Eq:energ_pow}
    \varepsilon(a_i, a_{i+r}) = -\varepsilon (2a_i - 1)(2a_{i+r}-1) r^{-\beta},
\end{equation}
where the coefficient $\beta$ in the exponent, generally speaking, is not equal to $\alpha$,
and $\alpha,\, \beta \gtrsim 1$.
\subsubsection{Entropy}

For the systems with week memory, $F_0 \ll 1$, the first approximation
for the normalized correlation function has the
form (see Ref.~\cite{MelUs}),
\begin{equation}\label{Eq:pow_weakcor}
    K(r) \simeq F(r) = F_0 r^{-\alpha},
\end{equation}
and we can calculate the entropy  per symbol Eq.~\eqref{Eq:EntropDiff}
for the $L$-block, $L \leqslant N$,
\begin{equation}
    h_L = \ln 2 - \frac{F_0^2}{2} \sum_{r=1}^L r^{-2\alpha}.
\end{equation}
For the values of $\alpha \gtrsim 1$, the sum $\sum_{r=1}^L \ldots$
can be approximated by the integral $\int_{1/2}^{L+1/2} dr \ldots$,
which leads to
\begin{equation}
    h_L = \ln 2 - \frac{F_0^2}{2(2\alpha - 1)} \left(\left(\frac{1}{2}\right)^{-(2\alpha-1)} - \left(L + \frac{1}{2}\right)^{-(2\alpha-1)}\right).
\end{equation}
Analogously, we obtain the approximate value of the block entropy,
\begin{eqnarray}
    &&H_{L} = \sum_{r=0}^{L} h_r = (L) \ln 2 \nonumber \\
        &&- (L-1)\frac{F_0^2}{2(2\alpha-1)}\left(\frac{1}{2}\right)^{-(2\alpha-1)} \\
        &&- \frac{F_0^2}{2(2\alpha-1)}\frac{(L)^{2-2\alpha} - 1}{2 - 2\alpha},\nonumber
\end{eqnarray}
or, for the case $L \gg 1$,
\begin{eqnarray}\label{Eq:pow_entr}
    H_L = L \left(\ln 2 -\frac{F_0^2}{2(2\alpha-1)}\left(\frac{1}{2}\right)^{-(2\alpha-1)}\right).
\end{eqnarray}
\subsubsection{Energy}

The average energy of the block of length $L$  we calculate
averaging Eq.~\eqref{Eq:energ_pow} over all the words $a_1^L$:

\begin{eqnarray*}
   && E_{L} = \sum_{a_1^L} P(a_1^L) E(a_1^L) \\
    &&= -\varepsilon \sum_{a_1^L} P(a_1^L) \sum_{i \neq j} (2a_i - 1)(2a_j-1) |i-j|^{-\beta} \\
    &&= -\varepsilon \sum_{i \neq j} |i-j|^{-\beta} \sum_{a_1^L} P(a_1^L) (2a_i - 1)(2a_j-1).
\end{eqnarray*}
Taking into account that the last sum over $a_1^L$ is just a
normalized correlation function $K(r)=C(r)/C(0)$ and using Eq.~\eqref{Eq:pow_weakcor}, we arrive at the following expression for
the average $L$-block  energy:
\begin{eqnarray*}
   && E_{L} = -2\varepsilon \sum_{i < j} (j-i)^{-\beta} K(j-i)\\
        &&= -2\varepsilon \sum_{r=1}^{L-1} (L-r) r^{-\beta/T} K(r)\\
      &&= -2\varepsilon F_0 \sum_{r=1}^{L-1} (L-r) r^{-(\alpha + \beta)}.
\end{eqnarray*}
For $L \gg 1$, approximating the sum with integral, we get
\begin{equation}
    E_{L} = -2^{\alpha+\beta-1} \varepsilon F_0 \left( \frac{2L}{\alpha+\beta-1} - \frac{1}{\alpha+\beta-2} \right),
\end{equation}
and, for $L \gg 1/(\alpha + \beta - 2)$,
\begin{equation}\label{Eq:pow_E}
    E_{L} = - \frac{2^{\alpha+\beta}}{\alpha+\beta-1} \varepsilon F_0 L.
\end{equation}
%
\subsubsection{Temperature}
At last, using Eqs.~\eqref{Eq:pow_entr} and \eqref{Eq:pow_E}, we can
obtain the desired estimation for the temperature of the Markov
chain with the power-law memory function,
\begin{equation}
   \frac{1}{\tau} = \frac{\varepsilon}{T} = \varepsilon \frac{dH_{L}}{dF_0} \Big/ \frac{dE_{L}}{dF_0} = 2^{(\alpha - \beta - 1)} \frac{\alpha + \beta - 1}{2\alpha - 1} F_0.
\end{equation}
Note that, for the special case $\alpha = \beta$, it is very simple:
\begin{equation}
   \frac{1}{\tau} = \frac{F_0}{2}.
\end{equation}
%

\subsection{Weakly correlated one-parametric Markov chains}
We have considered two models of Markov chains
for which we can introduce the information temperature. It seems
that the obtained results  allow us to make some general conclusions
on the possibility to introduce the temperature of the Markov chains
without making any assumptions about concrete model of their memory functions.

Fist of all, we suppose that the
Markov chain under study is one-parametric. This means that the
conditional probability function
$P(a_i=\alpha|a_{i-N},\ldots,a_{i-2},a_{i-1};\lambda)$ depends on some
parameter $\lambda$  proportional to the
amplitude of the correlation function. Without loss of generality,
we can suppose that this parameter belongs to the interval
$(\lambda_{min}, \lambda_{max})$. For the smallest value of $\lambda =
\lambda_{min}$, the function
$P(a_i=\alpha|a_{i-N},\ldots,a_{i-2},a_{i-1};\lambda)$ does not depend
on $a_{i-N},\ldots,a_{i-2},a_{i-1}$, i.e., it describes uncorrelated
sequence. When the parameter $\lambda$ tends to its maximum value, $\lambda
\rightarrow \lambda_{max}$, the conditional probability tends to unity if
all arguments $\alpha, a_{i-N},\ldots,a_{i-2},a_{i-1}$ coincide.
Note that the CPDF $P(.|.)$ of both considered above chains with memory functions
Eqs.~\eqref{1} and \eqref{Eq:pow}, satisfy this condition.
If the CPDF satisfies
the above condition, then we can introduce the information temperature, and the interval $(\lambda_{min}, \lambda_{max})$ can be mapped into the temperature interval $\tau \in(0,\infty)$.

Another example confirming
this conclusion, which we do not present here, is a random sequence
with correlation functions of exponential form, $K(r) = K_0 \exp
{(-\alpha r)}$, with constant value of the parameter $\alpha$.

For multi-parametric weakly correlated high-order additive Markov chains, the information temperature can be also introduced if one can distinguish, among others, the obvious parameter that determines the amplitude of the correlation function.

In the end of this section, we calculate the temperature for the literature texts considered
in Refs.~\cite{UYaKM, MUYaG}.
From the power long-range behaviour of memory function $F(r)=F_0
r^{-\alpha}$, we obtain the temperature, $\tau(F_0) =
2/F_0$.

\begin{center}
\begin{tabular}{||c|c|c||}
 \hline
 Text &  $F_0$ & $\tau(F_0)$ \\
 \hline
 Oliver Twist (Ch. Dickens)  & 0.1 & 20 \\
 \hline
 War and peace (L. Tolstoy)  & 0.1 & 20 \\
 \hline
 Tora  & 0.125 & 16 \\
 \hline
 Master and Margarita (M. Bulgakov)  & 0.2 & 10 \\
 \hline
 Don Quixote (M. Cervantes)  & 0.33 & 6 \\
 \hline
 Oblomov (I. Goncharov)  & 1.0 & 2 \\
 \hline
 Koran  & 1.4 & 1.4 \\
 \hline
\end{tabular}
\end{center}

%
\section{Conclusion and discussion}\label{Discussion}

\textbf{What is done.} In this paper, we suggest two approaches for introducing the information temperature of stationary ergodic random binary sequences.

i. Using the correspondence of the Markov chains with the two-sided
random sequences and comparing the two-sided sequences with their physical analogs
-- the Ising chains at given temperatures, we introduce the
information temperature for the Markov chain of the first and second orders.

ii. For the Markov chains of the $N$-th order with $N>3$, the entropy-based approach  described in Sections~\ref{2a},~\ref{3a}, and \ref{n3} appears to be more convenient. For the simplest case of Markov chains with the step-wise memory functions with $N=1,2$, both approaches give the same results.

iii. We also introduced the information temperature for the random weakly correlated
one-parametric Markov chains with step-wise and power memory functions. The results for the Markov chains
with step-wise memory functions at $\mu \ll 1$ coincide with ones obtained for $N=1,\,2,\,3$ within the second approach.

\textbf{Discussion.}
One of the main problems that we had to solve in this work was to determine the length of the
subsequence/word for which we introduce the information temperature. For the Markov chain of the $N$-th
order, it was chosen equal to $N + 1$. The reason for
this was the statement that all the statistical properties
of the chain under consideration are determined by the
statistics of the $(N+1)$-words. Are there any other ways to
determine their optimal length? One of the reasonable
answers is to study words which length tends to infinity.
Obviously, the temperature, as the macroscopic parameter
of the sequence, should go to some constant value as the
block length $L$ tends to infinity. This allows one to
consider this limiting value as a size-invariant property.
In Appendix C, we show that the suggested method  actually gives a constant value of $T$ at $L \rightarrow \infty$ for every
system.

It is necessary to emphasize some differences
between the notion of the ``information temperature'' considered here
from the classical thermodynamical one. In statistical physics, the
temperature is introduced via consideration of the thermodynamic
equilibrium between weakly interacting systems. What is an analogue
of such equilibrium for random sequences?  In the following
research, one have to explain this point. Now we can only stress that the temperature can be considered as an important macroscopic characteristics of a random sequence or text. We can also
indicate here the uniformity of distribution of the information temperature along
the random \emph{stationary} chain.

One of the important future development of this research is to consider symbolic random sequences with arbitrary (not binary) finite state space. It will be expedient also to take into account the
shifted average value of the elements of the Markov chain (with $\overline{a}\neq 1/2$) which
would be similar to the influence of an external magnetic field in
the Ising model. Deserves attention the study of information temperature of various RNA and DNA viruses, including various strains of SARS-CoV-2, and to compare the obtained results with the clinical manifestations of viruses. It is of interest to explore the information meaning of the
introduced temperature. In particular, a question arises: can the temperature
characterize the academic level of the text, or serve as an
indicator of the brain activity of the author? D. Tononi ~\cite{Tononi} indicates the possibility to characterize a neural network by some $\Phi$ function.
This function is not well defined and is repeatedly criticized. Can a neural network be characterized by the temperature of the text that the neural network is able to write?

\begin{acknowledgments}
We thank Drs. S.~S.~Apostolov, V.~E.~Vekslerchik, and Z.~A.~Maizelis for useful
discussions when doing the work. Initially, this work was supposed to be carried out with a support from the National Research Foundation of Ukraine, Project No. 2021.01/0016
``Mutual long-range correlations, memory functions, entropy, and information temperature of nucleotide sequences of RNA viruses as indicators of danger to Humans.''
However, the war in Ukraine led to the suspension of funding for this Project.
\end{acknowledgments}
\appendix 

\section{Block entropy}

Consider the block entropy of $L$-words with $L<N$,
\begin{equation}
H_L = - \sum_{w_L} P(w_L) \log P(w_L).
\end{equation}
Here $w_L$ is the concise notation for a subsequence ($L$-word,
tuple) of symbols, $w_L=a_1^L=a_1,...,a_L$. The entropy $H_{L+1}$ of
the concatenated word $w_{L+1}=(w_L, a)$ is

\begin{eqnarray}
H_{L+1} &= &- \sum_{w_L, a} P(w_L, a) \log P(w_L, a)\notag  \\
& =&
-\sum_{w_L} \sum_a P(w_L)P(a|w_L)  \\
  && \times\left( \log P(w_L)+ \log P(a|w_L)\right)\notag  \\
& =& H_L - \sum_{w_L} P(w_L) \sum_a P(a|w_L) \log P(a|w_L). \notag
\end{eqnarray}
Here the averaging $\sum_{w_L, a}$ is done over all symbols of the
words $w_{N+1}=a_1^{N+1}=a_1,...,a_{N},a$. So, the conditional
entropy (per symbol) $h_L \equiv H_{L+1} - H_L$ is
\begin{equation}
\label{Eq:sL_gen} h_L = - \sum_{w_L} P(w_L) \sum_a P(a|w_L) \log
P(a|w_L).
\end{equation}

If the length $L$ of word exceeds the memory length $N$, $L>N$, then
the conditional probability $P(a|w_L)$ is defined by previous (with
respect to symbol $a$) $N$ symbols, $N$-length part of the $L$-word only, $P(a|w_L) =
P(a|w_N)$. Consider each $L$-word in Eq.~\eqref{Eq:sL_gen} as a
combination of $(L-N)$- and $N$-words:
\begin{widetext}
\begin{eqnarray*}
h_L &=& - \sum_{w_{L-N}} \sum_{w_N} P(w_{L-N},w_N) \sum_a
P(a|w_{L-N},w_N) \log P(a|w_{L-N},w_N)\\ & =& \notag -
\sum_{w_{L-N}} \sum_{w_N} P(w_{L-N},w_N) \sum_a P(a|w_N) \log
P(a|w_N) \\ &=& - \sum_{w_N} \sum_a P(a|w_N) \log P(a|w_N)
\sum_{w_{L-N}} P(w_{L-N},w_N) \\ &=& \notag - \sum_{w_N} P(w_N)
\sum_a P(a|w_N) \log P(a|w_N) \equiv h_N.
\end{eqnarray*}
\end{widetext}
Thus, the differential entropy becomes constant at $L > N$,
\begin{equation}
\label{Eq:s_def} h_{L>N} = - \sum_{w_N} P(w_N) \sum_a P(a|w_N) \log
P(a|w_N) = h.
\end{equation}

Then, for $L \gg N$, the block entropy $H_L$ becomes a linear function of $L$,
\begin{equation}
H_{L \gg N} \approx hL.
\end{equation}

\section{Third order chains}

To calculate the information temperature of the Markov chain of the third
order we should use the same method as is presented in Sections
\ref{2a} and \ref{3a}. Using Eqs.~\eqref{1} -- \eqref{15} for $N=3$,
we find the probabilities of 3-symbol words occurring. Applying Eq.~\eqref{15}  for probabilities of 4-symbol words and using the found conditional probabilities for 3-words, we obtain the joint probabilities for all
possible symbol combinations of four-symbol blocks. They are as
followed:
\begin{eqnarray}
 P_{0000}=P_{1111}=\frac{3+8 \mu+4 \mu^2}{16(3-4 \mu)},
\end{eqnarray}  
\begin{eqnarray}
P_{0001}&=&P_{0010}=P_{0100}=P_{0111}=P_{1000} \nonumber \\
&=& P_{1011}=P_{1101}=P_{1110} \nonumber \\
&=&\frac{3-4 \mu-4 \mu^2}{16(3-4 \mu)},
\end{eqnarray}
\begin{eqnarray}
&&P_{0011}=P_{0101}=P_{0110}=P_{1001}\nonumber \\
&=&P_{1010} =P_{1100}=\frac{3-8\mu+4 \mu^2}{16(3-4 \mu)}. \nonumber \\
\end{eqnarray}
The entropy and energy are equal to
\begin{widetext}
\begin{eqnarray*}
&&H_4= \frac{1}{8 (4 \mu -3)}\left[ (2 \mu +3)\left[(2 \mu +1) \log
\left(\frac{4 \mu ^2+8 \mu +3}{48-64 \mu }\right)-(8 \mu -4) \log
\left(\frac{4 \mu ^2+4 \mu -3}{64 \mu -48}\right)\right] \right.\\
&& \left. +3 \left(4 \mu ^2-8 \mu +3\right) \log \left(\frac{4 \mu
^2-8 \mu +3}{48-64 \mu }\right)\right],
\end{eqnarray*}
\end{widetext}
%
%
%
\begin{equation}
E_4=\frac{2 \mu  \left(3 \varepsilon _1+2 \varepsilon _2+\varepsilon
_3\right)}{4 \mu -3}.
\end{equation}
Here we have introduced energies of interaction of the third order
$\varepsilon_3$, $\varepsilon_{\uparrow \bullet  \bullet\uparrow
}=\varepsilon_{\downarrow \bullet \bullet
\downarrow}=-\varepsilon_3$, $\varepsilon_{\uparrow \bullet \bullet
\downarrow }=\varepsilon_{\downarrow \bullet \bullet
\uparrow}=\varepsilon_3.$

\section{Energy and temperature of long fragments of the Markov sequences}\label{App:L_infty}

For arbitrary energy function,
\begin{equation}
\varepsilon(a_i, a_{i+r}) = -(2a_i - 1)(2a_{i+r} - 1)\varepsilon(r),
\end{equation}
using the definitions $K(r) = C(r)/C(0)$ and
$C(r)=\overline{(a_i-\overline{a})(a_{i+r}-\overline{a})}$, one can
get
\begin{equation}
E_{L} = -2 \sum_{r=1}^{L-1} (L-r) \varepsilon(r) K(r),
\end{equation}
which becomes linear in $L$ at $L \gg N$:
\begin{equation}
\label{Eq:E_def} E_{L \gg N} \approx EL, \,\,\,\, E = -2
\sum_{r=1}^\infty \varepsilon(r) K(r).
\end{equation}
%

Thus, at $L \rightarrow \infty$ the temperature goes to constant
value defined by the derivations,
\begin{equation}
\frac{1}{\tau} = \frac{\varepsilon}{T} = \varepsilon
\frac{dh}{d\lambda}\Big/ \frac{dE}{d\lambda},
\end{equation}
where the functions $h$ and $E$ are defined by Eqs.~\eqref{Eq:s_def}
and~\eqref{Eq:E_def}, and $\lambda$ is some parameter of the CPDF.
%

\end{document}